\documentclass[conference]{IEEEtran}
\IEEEoverridecommandlockouts
\pdfoutput=0
\usepackage[utf8]{inputenc} 
\usepackage[T1]{fontenc}
\usepackage{url}
\usepackage{ifthen}
\usepackage{cite}
\usepackage{verbatim} 
\usepackage{hyperref}
\usepackage[hyphenbreaks]{breakurl}
\usepackage[cmex10]{amsmath} % Use the [cmex10] option to ensure complicance
\usepackage{makecell}
\usepackage{textcase}
\usepackage[tablename=Table]{caption}
\usepackage{blindtext}
\usepackage{floatrow}
\usepackage{soul}
\usepackage{gensymb}
\usepackage{cite}
\usepackage{amsmath,amssymb,amsfonts}
\usepackage{textcomp}
\usepackage{graphicx}
\usepackage{amssymb}
\usepackage{amsfonts}
\usepackage{amsmath}
\usepackage{epsfig}
\usepackage{color}
\usepackage{fancybox}
\usepackage{textcomp}
\usepackage{multirow}
\usepackage{setspa ce}
\usepackage{psfrag}
\usepackage{booktabs}
\usepackage{float}
\usepackage{algorithm}
\usepackage{algpseudocode}
\usepackage{mathtools, nccmath, bigints, amsfonts}
\usepackage{subfigure}
\usepackage{caption}
\usepackage{mathrsfs}								% 支持花体（大写）

\usepackage{placeins}

\newfloatcommand{capbtabbox}{table}[][0.4\textwidth]

    \def\Complex{{\rm\rule[.23ex]{.03em}{1.1ex}\kern-.3em{C}}}

    \newcommand{\be}{\begin{equation}} \newcommand{\ee}{\end{equation}}
    \newcommand{\bea}{\begin{eqnarray}} \newcommand{\eea}{\end{eqnarray}}
    \newcommand{\benum}{\begin{enumerate}} \newcommand{\eenum}{\end{enumerate}}

        \newcommand{\qb}{{\bf b}}

        \newcommand{\qh}{{\bf h}}

        \newcommand{\qn}{{\bf n}}

        \newcommand{\qr}{{\bf r}}
        \newcommand{\qs}{{\bf s}}

        \newcommand{\qv}{{\bf v}}
        \newcommand{\qw}{{\bf w}}
        \newcommand{\qx}{{\bf x}}
        \newcommand{\qy}{{\bf y}}
        \newcommand{\qz}{{\bf z}}

        \newcommand{\qF}{{\bf F}}
        \newcommand{\qG}{{\bf G}}
        \newcommand{\qH}{{\bf H}}
        \newcommand{\qI}{{\bf I}}

        \newcommand{\qR}{{\bf R}}

        \newcommand{\qW}{{\bf W}}
        \newcommand{\qX}{{\bf X}}
        \newcommand{\qY}{{\bf Y}}

        \newcommand{\qDelta}{{\boldsymbol \Delta}}

        \newcommand{\qSigma}{{\boldsymbol \Sigma}}

        \newcommand{\qmu}{{\boldsymbol \mu}}

\def\BibTeX{{\rm B\kern-.05em{\sc i\kern-.025em b}\kern-.08em
	T\kern-.1667em\lower.7ex\hbox{E}\kern-.125emX}}
\begin{document}

\title{Untrained Neural Network based Bayesian Detector for OTFS Modulation Systems} 

\author{%
	\IEEEauthorblockN{
		Hao Chang\IEEEauthorrefmark{1},
		Alva Kosasih\IEEEauthorrefmark{1},
            Wibowo Hardjawana\IEEEauthorrefmark{1}, 
		Xinwei Qu\IEEEauthorrefmark{1},
		% Vera Miloslavskaya\IEEEauthorrefmark{1},
		% Victor Andrean\IEEEauthorrefmark{2},
		% Vincent Onasis\IEEEauthorrefmark{1},
		and Branka Vucetic\IEEEauthorrefmark{1}\\
	}
	\IEEEauthorblockA{\IEEEauthorrefmark{1}%
		Centre of Excellence in IoT and Telecommunications, The University of Sydney, Sydney, Australia.  \\
		\{hao.chang,alva.kosasih,wibowo.hardjawana,xinwei.qu,branka.vucetic\}@sydney.edu.au}
	
	%\IEEEauthorblockA{\IEEEauthorrefmark{2}%
		% Mobilizing Information Technology Lab., National Taiwan University of Science and Technology, Taipei, Taiwan. }
}

\maketitle

%%%%%%
%% Abstract: 
%% If your paper is eligible for the student paper award, please add
%% the comment "THIS PAPER IS ELIGIBLE FOR THE STUDENT PAPER
%% AWARD." as a first line in the abstract. 
%% For the final version of the accepted paper, please do not forget
%% to remove this comment!
%%
\begin{abstract}
   The orthogonal time frequency space (OTFS) symbol detector design for high mobility communication scenarios has received numerous attention lately. Current state-of-the-art OTFS detectors mainly can be divided into two categories; iterative and training-based deep neural network (DNN) detectors. Many practical iterative detectors rely on minimum-mean-square-error (MMSE) denoiser to get the initial symbol estimates. However, their computational complexity increases exponentially with the number of detected symbols. Training-based DNN detectors typically suffer from dependency on the availability of large computation resources and the fidelity of synthetic datasets for the training phase, which are both costly.
   In this paper, we propose an untrained DNN based on the deep image prior (DIP) and decoder architecture, referred to as D-DIP that replaces the MMSE denoiser in the iterative detector. DIP is a type of DNN that requires no training, which makes it beneficial in OTFS detector design. Then we propose to combine the D-DIP denoiser with the Bayesian parallel interference cancellation (BPIC) detector to perform iterative symbol detection, referred to as D-DIP-BPIC. Our simulation results show that the symbol error rate (SER) performance of the proposed D-DIP-BPIC detector outperforms practical state-of-the-art detectors by 0.5 dB and retains low computational complexity.
\end{abstract}


%% The paper must be self-contained. However, if you are referring to
%% a full version for checking certain proofs, please provide the
%% publically accessible location below.  If the paper is completely
%% self-contained, you can remove the following line from your
%% submission.+

\section{Introduction}

The future mobile system will support various high-mobility scenarios (e.g., unmanned aerial vehicles and autonomous cars) with strict mobility requirements \cite{hadani2017orthogonal}.
However, current orthogonal frequency division multiplexing (OFDM)\cite{jiang2010channel} is not suitable for these scenarios due to the high inter-carrier interference (ICI) caused by a large number of high-mobility moving reflectors.
The orthogonal time frequency space (OTFS) modulation was proposed in \cite{hadani2017orthogonal} to address this issue because it allows the tracking of ICI during the symbol estimation process.
Multiple OTFS symbol detectors \cite{raviteja2018interference,khumalo2016fixed,9569353,9492800,long2021low,singh2020low,9518377,9900413} have been investigated in current literature.

%Linear detector, iterative detector

Several iterative detectors have been proposed in OTFS systems, e.g., message passing (MP)\cite{raviteja2018interference}, approximate message passing (AMP)\cite{khumalo2016fixed}, Bayesian parallel interference cancellation (BPIC) that uses minimum-mean-square-error (MMSE) denoiser \cite{9569353}, unitary approximate message passing (UAMP) \cite{9492800}, and expectation propagation (EP)\cite{long2021low} detectors.
These detectors provide a significant symbol error rate (SER) performance gain compared to that of the classical MMSE detector\cite{singh2020low}.
Unfortunately, when a large number of moving reflectors exist, MP and AMP suffer from performance degradation due to high ICI \cite{9569353}. 
The UAMP detector addresses this issue by performing singular value decomposition (SVD) that exploits the structure of the OTFS channel prior to executing AMP.
Similar performance in terms of reliability and complexity to the UAMP detector has also been achieved by our proposed iterative MMSE-BPIC detector in \cite{9569353}. We combined an MMSE denoiser, the Bayesian concept, and parallel interference cancellation (PIC) to perform iterative symbol detection.
Unfortunately, their performance is still suboptimal in comparison with the EP OTFS detector \cite{long2021low}. 
EP uses the Bayesian concept and multivariate Gaussian distributions to approximate the mean and variance of posterior detected symbols iteratively from the observed received signals. 
The outperformance of the EP detector comes at the cost of high computational complexity in performing iterative matrix inversion operations.
%Note that although the detector proposed in \cite{9832663} referred to as the GEPNet can achieve a significantly better performance than the EP and thus it can be regarded as the current best symbol detector, the GEPNet detector has not been implemented in OTFS systems and suffers a high complexity problem as its predecessor EP due to performing matrix inversion in every iteration. 

%NN detector 

In addition to those iterative detectors, deep neural network (DNN) based approaches are widely used in symbol detector design. They can be divided into two categories; 1) Training-based DNN and 2) untrained DNN. The training-based DNN requires a large dataset to train the symbol detector prior to deployment. Recent examples of training-based DNN category are a 2-D convolutional neural network (CNN) based OTFS detector in \cite{9518377} 
%, our previously proposed GPICNet multiple-input multiple-output (MIMO) detector that uses graph neural network (GNN) to improve the mean and variance of the posterior probability symbol estimates of PIC detector \cite{9832663}
and also our recently proposed BPICNet OTFS detector in \cite{9900413} that integrates the MMSE denoiser, BPIC and DNN whereby the modified BPIC parameters are trained by using DNN. There are two major disadvantages for the training-based DNN approach; 1) dependency on the availability of large computation resources that necessitate substantial energy or CO2 consumptions and high cost for the training phase \cite{strubell-etal-2019-energy}; 2) the fidelity of synthetic training data, artificially generated due to high cost of acquiring real datasets, in the real environment \cite{AI}. For example, a high fidelity training dataset implies the distribution functions for all possible velocity of mobile reflectors is known beforehand, which is impossible.

%Unfortunately, the performance of trained DNN depends on the number of trainable parameters and the completeness of training datasets in representing various wireless environments. In practice, the cost of acquiring these large training datasets is prohibitively expensive.

The second category, untrained DNN, avoids the need for training datasets.
Deep image prior (DIP) proposed in \cite{ulyanov2018deep} has been widely used in image restoration as an untrained DNN approach.
The encoder-decoder architecture used in the original DIP shows excellent performance in image restoration tasks but the use of up to millions of trainable parameters results in high latency and thus still cannot be used for an OTFS detector that requires close to real-time processing time.
Recently, the authors in \cite{9488215} show that the decoder-only DIP offers similar performance as compared to an encoder-decoder DIP architecture when it is applied to Magnetic Resonance Imaging (MRI). The complexity of decoder-only DIP is significantly lower than the original encoder-decoder DIP, thus enhancing its potential use as a real-time OTFS detector. To date, no study has been conducted on untrained DNN based OTFS detectors.

In this paper, we propose to use untrained DNN with BPIC to perform iterative symbol detection. Specifically, we use DIP with a decoder-only architecture, referred to as D-DIP to act as a denoiser and to provide the initial symbol estimates for the BPIC detector. We choose BPIC here in order to keep low computational complexity for the OTFS receiver. 
We first describe a single-input single-output (SISO) OTFS system model consisting of the transmitter, channel and receiver.
We then provide a review of the MMSE-BPIC detector in \cite{Kosasih_2021,9569353} that uses the MMSE denoiser to obtain the initial symbol estimates. 
%that embeds parallel interference cancellation (PIC) and the Bayesian concept to iteratively estimate the mean and variance of the transmitted symbols, decision statistics combining (DSC) concept to weight the soft symbol estimates based on the previous and current iterations, and the MMSE filter to obtain the initial symbol estimates. 
%Moreover, we provide evidence of the sensitivity of the initial symbol estimates for the BPIC detector.
Instead of using MMSE, we propose a high-performance D-DIP denoiser to calculate the initial symbol estimates inputted to the BPIC. 
We then explain our proposed D-DIP in detail and also provide computational complexity and performance comparisons to other schemes. Simulation results indicate an average of approximately 0.5 dB SER out-performance as compared to other practical schemes in the literature.

%Simulation results show that the proposed D-DIP-BPIC OTFS detector can achieve an approximate 0.5 dB performance gain over UAMP and MMSE-BPIC.

The main contribution of this paper is the first to propose a combination of a decoder-only DIP denoiser and the BPIC OTFS detector. The proposed denoiser 1) provides better initial symbol estimates for the BPIC detector and 2) has lower computational complexity than the MMSE denoiser.  This leads to the proposed scheme having the closest SER performance to the EP scheme as compared to other schemes, achieved with much lower computational complexity (approximately 15 times less complex than the EP).

%\addtolength{\topmargin}{0.05in}
\begin{figure*}
	\centering
	\includegraphics[width=0.9\textwidth]{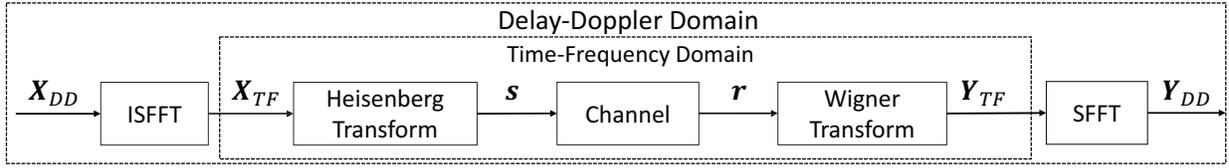}
	\caption{The system model of OTFS modulation scheme}
	\label{fig:sys-mod-general}
\end{figure*}

{\bf Notations}: $a$, $\textbf{a}$ and $\textbf{A}$ denote scalar, vector, and matrix respectively. $\mathbb{C}^{M\times N}$ denotes the set of $M\times N$ dimensional complex matrices. We use $\qI_N$, $\textbf{F}_N$, and $\textbf{F}_N^{\qH}$ to represent an $N$-dimensional identity matrix, $N$-points discrete Fourier Transform (DFT) matrix, and $N$-points inverse discrete Fourier transform (IDFT) matrix. 
%$(\cdot)^T$, $(\cdot)^{\qH}$, $(\cdot)^*$, and $[\cdot]_M$ represent the transpose, Hermitian,  conjugate, and mod-$M$ operations. 
$(\cdot)^T$ represents the transpose operation.
We define $\textbf{a} = {\sf vec}(\textbf{A})$ as the column-wise vectorization of matrix $\textbf{A}$ and $\textbf{A} = {\sf vec}^{-1}(\textbf{a})$ denotes the vector elements folded back into a matrix.
The Kronecker product is denoted as $\otimes$. 
$\lfloor \frac{a}{b} \rfloor$ represents the floor operation, and $[\cdot]_M$ represent the mod-$M$ operations.
% and $a\%b$ represent the modulo operation for scalar $a$ and $b$.
The Euclidean distance of vector $\qx$  is denoted as $\|\qx\|$. We use $\mathcal{N}(\qx:\qmu, \qSigma)$ to express the multivariate Gaussian distribution of a vector $\qx$ where $\qmu$ is the mean and $\qSigma$ is the covariance matrix. 
%$\mathbb{E}[\qx]$ is the mean of random vector $\qx$. $\sf diag(\qA)$ refers to set the non-diagonal elements of the matrix $\qA$ to zero and $\sf{tr}(\qA)$ denote the trace of $\qA$.

\section{OTFS System Model}

We consider an OTFS system, as illustrated in Fig. \ref{fig:sys-mod-general}. In the following, we explain the details of the OTFS transmitter, channel and receiver.

\subsection{OTFS Transmitter}

%In the transmitter side, an $M$-ary quadrature amplitude modulation ($M$-QAM) symbol $\qX_{\rm DD}$ is mapped from a binary sequence in the delay Dopper (DD) domain. The constellation set of the $M$-QAM symbols is denoted as $\Omega$. 
In the transmitter side, $MN$ information symbols $\qX_{\rm DD} \in \mathbb{C}^{M \times  N}$ from a modulation alphabet of size $Q \mathbb{A}=\{a_1,\cdots,a_Q\}$ are allocated to an $M \times N$ grids in the delay-Doppler (DD) domain, where $M$ and $N$ represent the number of subcarriers and time slots used, respectively.
%$m=0,\cdots,M-1$ and $n=0,\cdots,N-1,$ are the indices of discretized delay and Doppler shifts, respectively. 
As illustrated in Fig. \ref{fig:sys-mod-general}, the DD domain symbols are transformed into the time-frequency (TF) domain by using the inverse symplectic finite Fourier transform (ISFFT)\cite{hadani2017orthogonal}. Here, the TF domain is discretized to $M$ by $N$ grids with uniform intervals $\Delta f$ (Hz) and $T_s=1/\Delta f$ (seconds), respectively. Therefore, the sampling time is $T_s /M$.
The TF domain sample $\qX_{\rm TF} \in \mathbb{C}^{M\times N}$ is an OTFS frame, which occupies the bandwidth of $M\Delta f$ and the duration of $NT_s$, is given as
\begin{equation}
	\qX_{\rm TF} = \qF_M\qX_{\rm DD}\qF_N^{\qH},
	\label{eq:sysmod-tx-isfft}
\end{equation}
where $\qF_M \in \mathbb{C}^{M\times M}$ and $\qF_N^{\qH} \in \mathbb{C}^{N\times N}$ are $M$-points DFT and $N$-points IDFT matrices, and the $(p,q)$-th entries of them are $(\frac{1}{\sqrt{M}}e^{-j2\pi pq/M})_{p,q=0,\cdots,M-1}$ and $(\frac{1}{\sqrt{N}}e^{j2\pi pq/N})_{p,q=0,\cdots,N-1}$, respectively. The $(m,n)$-th entries $X_{\rm TF}[m,n]$ of $\qX_{\rm TF}$ is written as
\begin{equation}
	X_{\rm TF}[m,n] = \frac{1}{\sqrt{MN}} \sum_{k=0}^{N-1} \sum_{l=0}^{M-1} 	X_{\rm DD}[k,l]e^{j2\pi(\frac{nk}{N}-\frac{ml}{M})},
	\label{x_tf[m,n]}
\end{equation}
where $X_{\rm DD}[k,l]$ represents the $(k,l)$-th entries of $\qX_{\rm DD}$ for $k=0,\cdots,M-1, l=0,\cdots,N-1$.
The (discrete) Heisenberg transform\cite{hadani2017orthogonal} is then applied to generate the time domain transmitted signal by using \eqref{eq:sysmod-tx-isfft} and Kronecker product rule\footnote{A matrix multiplication is often expressed by using vectorization with the Kronecker product. That is, ${\sf vec}(ABC) = (C^T \otimes A){\sf vec}(B)$}, the vector form of the transmitted signal can be written as
\begin{equation}\label{eq:sysddmod-tx-heisenberg}
    \qs = {\sf vec}(\qG_{\rm tx}\qF^{\qH}_M\qX_{\rm TF}) = (\qF^{\qH}_N\otimes \qG_{\rm tx})\qx_{\rm DD},    
	%&={\sf vec}(\qG_{\rm tx}\qF^{\qH}_L    \qF_L\qX_{\rm DD}\qF_K^{\qH})  \notag \\ 
	%& = {\sf vec}(\qG_{\rm tx}\qX_{\rm DD}\qF_N^{\qH}) \notag\\
	% = (\qF^{\qH}_N\otimes \qG_{\rm tx})\qx_{\rm DD}
\end{equation}
where $\qG_{\rm tx}$ is the pulse-shaping waveform, and we consider the rectangular waveform with a duration of $T_s$ that leads to $\qG_{\rm tx}=\qI_M$ \cite{8516353},
$\qx_{\rm DD} = {\sf vec}(\qX_{\rm DD})$, and $\qx_{\rm DD}=[x_{\rm DD}(0),\cdots,x_{\rm DD}(MN-1)]^T$. $\qs \in \mathbb{C}^{MN\times 1}$ is the vector form of the transmitted signal, $\qs=[s(0),\cdots,s(n),\cdots,s(MN-1)]^T$, $n=0,\cdots,MN-1$, and $s(n)$ can be written as
\begin{equation}
    s(n) = \frac{1}{\sqrt{N}} \sum_{k=0}^{N-1} e^{j2\pi \lfloor \frac{n}{M} \rfloor k/N} x_{\rm DD}([n]_M+kM).
\end{equation}
%\begin{equation}
%    \begin{split}
%    \qs(n) = \frac{1}{\sqrt{N}}x_{\rm DD}([n]_M))+ 
%    \frac{e^{j2\pi \lfloor \frac{n}{M}\rfloor/N}}{\sqrt{N}} x_{\rm %DD}([n]_M+M) + \\ 
%    \cdots+ \frac{e^{j2\pi \lfloor \frac{n}{M} \rfloor %(N-1)/N}}{\sqrt{N}} x_{\rm DD}([n]_M+(N-1)M),    
%    \end{split}
%\end{equation}
We insert the cyclic prefix (CP) at the beginning of each OTFS frame, the length of CP is the same as the index of maximum delay $l_{max}$. Thus, the time duration after adding CP is $NT_s + N_{\rm cp}\frac{T_s}{M}$, where $ N_{\rm cp}=l_{max}$. 
%\blue{I don't think it is necessary to include the below expression regarding CP.}
After adding CP, $\qs = [s(MN-N_{\rm cp}+1),s(MN-N_{\rm cp}+2),\cdots,s(MN-1),s(0),\cdots,s(n),\cdots,s(MN-1)]^T$, and $\qs$ is transmitted through a time-varying channel.

\subsection{OTFS Wireless Channel}
The OTFS wireless channel is a time-varying multipath channel, represented by the impulse responses in the DD domain,
\begin{equation}\label{channel}
	h(\tau, v) = \sum_{i=1}^P h_i \delta(\tau - \tau_i)\delta(v - v_i)
\end{equation}
where $\delta(\cdot)$ is the Dirac delta function, $h_i \sim \mathcal{N}(0, 1/P)$ denotes the $i$-th path gain, and $P$ is the total number of paths. Each of the paths represents a channel between a moving reflector/transmitter and a receiver with a different delay $(\tau_i)$ and/or Doppler $(v_i)$ characteristics.
The delay and Doppler shifts are given as $\tau_i = l_i\frac{T_s}{M}$ and$\quad v_i = k_i\frac{\Delta f}{N}$, respectively. The ICI depends on the delay and Doppler of the channel as illustrated in \cite{8516353}. Here, for every path, the randomly selected integers $l_i \in [0,  l_{max}]$ and $k_i \in [-k_{max}, k_{max}]$ denote the indices of the delay and Doppler shifts,  where $l_{max}$ and $k_{max}$ are the indices of the maximum delay and maximum Doppler shifts among all channel paths. Note for every path, the combination of the $l_i$ and $k_i$ are different.
For our wireless channel, we assume $l_{max} \leq M-1$ and $k_{max} \leq \lfloor \frac{N}{2} \rfloor $, implying maximum channel delay and Doppler shifts of less than $T_s$ seconds and $\Delta f$ Hz, respectively.

\subsection{OTFS Receiver}
At the receiver side, the time domain received signal $r(t)$ is shown as \cite{hadani2017orthogonal}
\begin{equation}\label{r(t)}
    r(t) = \int \int h(\tau,v)s(t-\tau)e^{j2\pi v(t-\tau)} d\tau dv + w(t),
\end{equation}
where $s(t)$ is the time-domain received signal $\qs$, while $h(\tau,v)$ is the DD domain channel shown in \eqref{channel}.
The received signal $r(t)$ is then sampled at $t = \frac{n}{M \Delta f}$, where $n=0,\cdots,MN-1$. After discarding CP, the discrete received signal $r(n)$ is obtained from \eqref{channel} and \eqref{r(t)}, written as 
\begin{equation}\label{r(n)}
r(n) =  \sum^P_{i=1} h_i e^{j2\pi \frac{k_i(n-l_i)}{MN}} s([n - l_i]_{MN}) + w(n),
\end{equation} 
We then write \eqref{r(n)} in the vector form as
\begin{equation}
	\qr = \qH\qs + \qw,
	\label{eq:sysmod-r(n)}
\end{equation}
where $\qw$ is the complex independent and identically distributed (i.i.d.) white Gaussian noise that follows $\mathcal{N}(\bold{0}, \sigma^2_c \qI)$, $\sigma^2_c$ is the variance of the noise. $\qH = \sum_{i=1}^P h_i \qI_{MN}(l_i) \qDelta({k_i})$, $\qI_{MN}(l_i)$ denotes a $MN \times MN$ matrix obtained by circularly left shifting the columns of the identity matrix by $l_i$.
%for example when $l_i =1$,
%\[
%\qI_{KL}(1) = \begin{bmatrix}
%	0 & \cdots & 0 & 1\\
%	1 & \ddots & 0 & 0\\
%	\vdots & \ddots & \ddots & \vdots\\
%	0 & \cdots & 1 & 0\\
%\end{bmatrix}	.
%\]
$\qDelta$ is the $MN\times MN$ Doppler shift diagonal matrix, $\qDelta({k_i}) = {\sf diag}\left[e^{\frac{j2\pi k_i(0)}{MN}}, e^{\frac{j2\pi k_i(1)}{MN}}, \cdots, e^{\frac{j2\pi k_i(MN - 1)}{MN}}\right]$, and ${\sf diag}(\cdot)$ denotes a diagonalization operation on a vector.
Note that the matrices $\qI_{MN}(l_i)$ and $\qDelta({k_i})$ model the delay and Doppler shifts in \eqref{channel}, respectively. 
As shown in Fig. \ref{fig:sys-mod-general}, the TF domain received signal $\qY_{\rm TF} \in \mathbb{C}^{M\times N}$ is obtained by applying the Wigner transform \cite{8516353}, shown as,
\begin{equation}
	\qY_{\rm TF} = \qF_M\qG_{\rm rx}\qR,
	\label{eq:sysddmod-rx-wigner}
\end{equation}
where $\qR={\sf vec}^{-1}(\qr)$, $\qG_{\rm rx}$ is the rectangular waveform with a duration $T_s$ in the receiver, and $\qG_{\rm rx}=\qI_M$.
Then the DD domain received signal $\qY_{\rm DD} \in \mathbb{C}^{M\times N}$ is obtained by using the symplectic finite Fourier transform (SFFT), which is

%\begin{align}\label{eq:sysmod-rx-sfft}
%	\qY_{\rm DD}  &=\qF^{\qH}_M \qY_{\rm TF} \qF_N \notag \\   &=\qF^{\qH}_M \qF_M\qG_{\rm rx}\qR \qF_N  \notag\\
%	&= \qG_{\rm rx}\qR\qF_N.
%\end{align}
\begin{align}\label{eq:sysmod-rx-sfft}
	\qY_{\rm DD}  =\qF^{\qH}_M \qY_{\rm TF} \qF_N =\qF^{\qH}_M \qF_M\qG_{\rm rx}\qR \qF_N  = \qG_{\rm rx}\qR\qF_N.
\end{align}
By following the vectorization with Kronecker product rule, we can rewrite \eqref{eq:sysmod-rx-sfft} as
%\begin{align}
%	\qy_{\rm DD} &= {\sf vec}(\qY_{\rm DD}) \notag \\  
%	&= {\sf vec}(\qG_{\rm rx}\qR\qF_N) \notag \\  
%    &= (\qF_N \otimes \qG_{\rm rx})\qr.
%	\label{eq:sysmod-rx-vec}
%\end{align}
\begin{align}
	\qy_{\rm DD} = {\sf vec}(\qY_{\rm DD}) = {\sf vec}(\qG_{\rm rx}\qR\qF_N) = (\qF_N \otimes \qG_{\rm rx})\qr.
	\label{eq:sysmod-rx-vec}
\end{align}
By substituting \eqref{eq:sysddmod-tx-heisenberg} into \eqref{eq:sysmod-r(n)} and \eqref{eq:sysmod-rx-vec} we obtain
\begin{align}
	\qy_{\rm DD} = \qH_{\rm DD}\qx_{\rm DD} + \tilde{\qw},
	\label{sysddmod-y=hx+w}
\end{align}
where $\qH_{\rm DD}=(\qF_N \otimes \qG_{\rm rx})\qH(\qF^{\qH}_N \otimes \qG_{\rm tx})$ and $\tilde{\qw} =(\qF_N \otimes \qG_{\rm rx})\qw$ denote the effective channel and noise in the DD domain, respectively.  Here, $\tilde{\qw}$ is an i.i.d. Gaussian noise, since $\qF_N \otimes \qG_{\rm rx}$ is a unitary orthogonal matrix \cite{hadani2017orthogonal,8516353}.
For convenience, we transform complex-valued model in \eqref{sysddmod-y=hx+w} into real-valued model. Accordingly, $\qx= \begin{bmatrix}
	\Re(\qx_{\rm DD})  \ \Im(\qx_{\rm DD})
\end{bmatrix}^T$,
$\qy=
\begin{bmatrix}
	\Re(\qy_{\rm DD})  \ \Im(\qy_{\rm DD})
\end{bmatrix}^T$,
$\qn=
\begin{bmatrix}
	\Re(\tilde{\qw})  \ \Im(\tilde{\qw})
\end{bmatrix}^T$,
$\qH_{\rm eff}=
\begin{bmatrix}
	\Re(\qH_{\rm DD})  & -\Im(\qH_{\rm DD}) \\ \Im(\qH_{\rm DD}) & \Re(\qH_{\rm DD})
\end{bmatrix}$, $\Re (\cdot)$ and $\Im (\cdot)$ are the real and imaginary parts, respectively. Thus, the variance of $\qn$ is $\sigma^2 = \sigma^2_c /2$ and $\qx, \qy, \qn$ are vectors of size $2MN$ and $\qH_{\rm eff}$ is a matrix of size $2MN \times 2MN$. Then, we can rewrite \eqref{sysddmod-y=hx+w} as
\begin{align}
	\qy = \qH_{\rm eff} \qx + \qn.
	\label{real-y=hx+w}
\end{align}	
We assume $\qH_{\rm eff}$ is known at the detector side. For notation simplicity, we omit the subscript of $\qH_{\rm eff}$ in \eqref{real-y=hx+w} and just write it as $\qH$ in all subsequent sections. The signal-to-noise ratio (SNR) of the system is defined as $\rm SNR= 10log_{10} ( \frac{1}{\sigma^2_c}) dB$.

\section{MMSE-BPIC Detector}
\begin{figure}
	\centering
	\includegraphics[width=\textwidth]{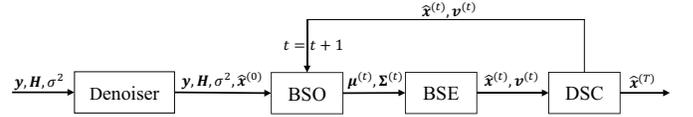}
	\caption{BPIC detector architecture}
	\label{BPIC structure}
\end{figure}

In this section, we briefly describe the BPIC detector that employs MMSE denoiser, recently proposed in \cite{Kosasih_2021}. 
%We then explain the problem with the initial symbol estimates process used by the BPIC detector. 
The structure of the BPIC detector is shown in Fig. \ref{BPIC structure}. It consists of four modules: Denoiser, Bayesian symbol observation (BSO), Bayesian symbol estimation (BSE), and decision statistics combining (DSC).
%\subsection{Bayesian Symbol Observation}

In the Denoiser module, the MMSE scheme is used to obtain the initial symbol estimates $\hat{\qx}^{(0)}$ in the first BPIC iteration \cite{Kosasih_2021} as shown in Fig. \ref{BPIC structure}. The MMSE denoiser can be expressed as
\begin{equation}\label{MMSE_init}
\hat{\qx}^{(0)} = \left(\qH^{T} \qH + \sigma^2\qI \right)^{-1}\qH^{T} \qy.
\end{equation}

In the BSO module, the matched filter based PIC scheme is used to detect the transmitted symbols, shown as
\begin{equation}\label{BSO_x}
\mu_q^{(t)} = \hat{x}_q^{(t-1)} + \frac{\qh_q^{T} \left( \qy-\qH \hat{\qx}^{(t-1)} \right)}{\| \qh_q \|^2},
\end{equation} 
where $\mu_q^{(t)}$ is the soft estimate of $q$-th symbol $x_q$ in iteration $t$, $\qh_q$ is the $q$-th column of matrix $\qH$. $\hat{\qx}^{(t-1)} =[\hat{x}_1^{(t-1)},\cdots,\hat{x}_q^{(t-1)},\cdots,\hat{x}_{2MN}^{(t-1)}]^T$ is the vector of the estimated symbol.
%and \[  \qx_{{\rm PIC} \backslash q}^{(t-1)}   = \left[x_{{\rm PIC}, 1}^{(t-1)}  , \dots, x_{{\rm PIC}, q-1}^{(t-1)} , 0, x_{{\rm PIC}, q+1}^{(t-1)}, \dots, x_{{\rm PIC}, MN}^{(t-1)} \right]^{\rm T} \] is the estimated symbols in the $(t-1)$-th iteration. 
The variance $ \Sigma_q^{(t)} $ of the $q$-th symbol estimate is derived in\cite{Kosasih_2021} as
\begin{align}\label{BSO_v}
\Sigma_q^{(t)} &   = \frac{1}{(\qh_q^T \qh_q)^2} \left(\sum_{j=1\atop{j\neq{q}}}^{MN}(\qh_q^T \qh_q)^2 v_j^{(t-1)}+(\qh_q^T \qh_q) \sigma^2 \right),
\end{align}
where $v_j^{(t-1)}$ is the $j$-th element in a vector of symbol estimates variance $\qv^{(t-1)}$ in iteration $t-1$ and $\qv^{(t-1)} = [v_1^{(t-1)},\cdots,v_q^{(t-1)},\cdots,v_{2MN}^{(t-1)}]^T$, we set $\qv^{(0)} = 0$ because we have no prior knowledge of the variance at the beginning.
Then the estimated symbol $\qmu^{(t)}=[\mu_1^{(t)},\cdots,\mu_q^{(t)},\cdots,\mu_{2MN}^{(t)}]^T$and variance $\qSigma^{(t)}=[\Sigma_1^{(t)},\cdots,\Sigma_q^{(t)},\cdots,\Sigma_{2MN}^{(t)}]^T$ are forwarded to the BSE module, as shown in Fig. \ref{BPIC structure}
%The approximated posterior functions,  $\hat{p}^{(t)}(x_q|\qy)={\cal{N}}\left( x_q,x^{(t)}_{{\rm PIC},q}; {\Sigma}^{(t)}_q \right), q=1, \dots, MN,$  are then forwarded to the BSE module, as shown in Fig. \ref{BPIC structure}.

%\subsection{Bayesian Symbol Estimator}
  
In the BSE module, we compute the Bayesian symbol estimates and the variance of the $q$-th symbol obtained from the BSO module. given as 

\begin{equation}\label{Bayesian_x}
\hat{x}_q^{(t)} =\mathbb{E} \left[x_q \Big| \mu_q^{(t)} ,\Sigma_q^{(t)} \right] =\sum_{a \in \Omega} a  \hat{p}^{(t)}{\left(x_q=a|\qy\right)}
\end{equation}
\begin{equation}\label{Bayesian_v}
v_q^{(t)}= \mathbb{E}  \left[ \left| x_q  - \mathbb{E} \left[x_q \Big| \mu_q^{(t)} ,\Sigma_q^{(t)} \right] \right|^{2} \right], 
\end{equation}
where $\hat{p}^{(t)}\left(x_q|\qy\right) = \mathcal{N}(x_q:\mu_q^{(t)}, \Sigma_q^{(t)}) $ is obtained from the BSO module and it is normalized so that  $\sum_{a\in \Omega}\hat{p}^{(t)}\left(x_q=a|\qy\right) =1$.
The outputs of the BSE module, $\hat{x}_q^{(t)} $and $v_q^{(t)}$  are then sent to the following DSC module.

%\subsection{Decision Statistics Combining}

The DSC module performs a linear combination of the symbol estimates in two consecutive iterations, shown as
\begin{equation}\label{DSC}
\hat{x}_q^{(t)} = \left( 1-\rho_q^{(t)} \right)  \hat{x}_q^{(t-1)}  +   \rho_q^{(t)}   \hat{x}_q^{(t)}
\end{equation}
\begin{equation}\label{DSC_Var}
 v_q^{(t)} = \left( 1-\rho_q^{(t)} \right)  v_q^{(t-1)}  +   \rho_q^{(t)}   v_q^{(t)}.
\end{equation}
The weighting coefficient is determined by maximizing the signal-to-interference-plus-noise-ratio  variance, given as 
\begin{equation}\label{DSC_coef}
\rho_q^{(t)} =  \frac{e_q^{(t-1)}}{e_q^{(t)}+e_q^{(t-1)}}, 
\end{equation}
where $e_q^{(t)}$ is defined as the instantaneous square error of the $q$-th symbol estimate, computed by using the MRC filter,
\begin{flalign}\label{DSC_error}
e_q^{(t)}  =  \left\|\frac{\qh_q^{T}}{\| \qh_q\|^2} \left(  \qy - \qH \hat{\qx}^{(t)} \right)\right\|^2.
\end{flalign}
The weighted symbol estimates $\hat{\qx}^{(t)}$  and their variance $\qv^{(t)}$ are then returned to the BSO module to continue the iteration. After $T$ iterations, $\hat{\qx}^{(T)}$ is taken as a vector of symbol estimates.

%\begin{figure}
%	\centering
%	\includegraphics[width=0.9\textwidth]{img/Init_compare.eps}
%	\caption{SER performance with different initialization scheme}
%	\label{init_comparation}
%\end{figure}

%\subsection{Different Initial symbol estimates for BPIC}

%The symbol estimates for the BPIC detector can be initialized differently.
%For example, when we do not have any prior information on the symbol estimates, we set  $\hat{\qx}^{(0)}=0$. This affects the symbol estimates accuracy as inaccurate initial symbol estimates leads to an estimation error that propagates over the iterations. To evaluate this, we simulate the two strategies of initial symbol estimates in the BPIC i.e., the MMSE and unknown initial symbol estimates. 
%Fig. \ref{init_comparation} shows the simulation results for the two different initial symbol estimates schemes with OTFS system configurations, $M=12,N=7,P=12,l_{max}=11,k_{max}=3$. 
%It is evident that the BPIC using MMSE denoiser significantly outperforms the one with unknown initial symbol estimates. This implies that the performance of the BPIC highly depends on the quality of initial symbol estimates.
\begin{figure*}
	\centering
	\includegraphics[width=0.85\textwidth]{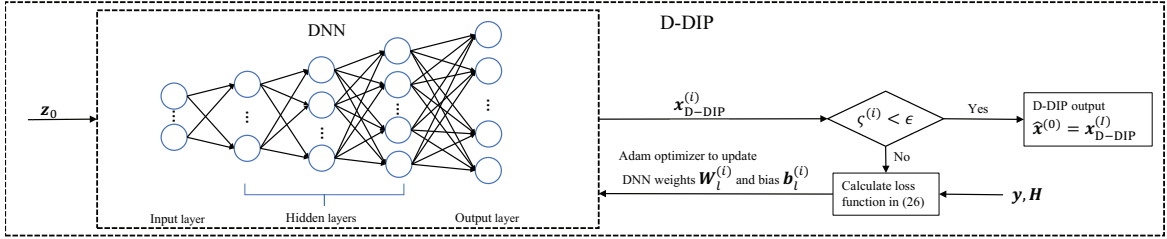}
	\caption{D-DIP structure}
	\label{fig:NN}
\end{figure*} 

\section{D-DIP denoiser For symbol estimation}\label{D-DIP}

In this section, we propose D-DIP to improve the initial symbol estimates performance of the BPIC detector, and the whole iterative process of D-DIP is shown in Fig. \ref{fig:NN}. 

The DNN used in D-DIP is classified as a fully connected decoder DNN that consists of $L=5$ fully connected layers.
Those layers can be broken down into an input layer, an output layer and three hidden layers with p1 = 4, p2 =
8, p3 = 16, p4 = 32, p5 = $2MN$ neurons, respectively.
We use a random vector $\qz_0$ drawn from a normal distribution $\mathcal{N}(\bold{0}, \bold{1})$ of size 4x1 as the input of the DNN first layer (i.e., input layer). $\qz_0$ is fixed during the D-DIP iterative process.

DNN output at iteration $i$ $\qx_{\rm {D-DIP}}^{(i)}$ is obtained by passing $\qz_0$ through 5 layers, shown as
\begin{equation}\label{NN_output}
   \qx_{\rm {D-DIP}}^{(i)} = c f_L^{(i)}(f_{L-1}^{(i)}(\cdots f_2^{(i)}(\qz_0))),
\end{equation}
where $c$ is a constant used to control the output range of the DNN and $f_l^{(i)}$ is the output of layer $l$ at iteration $i$,
\begin{equation}\label{Layer_output}
     f_l^{(i)} = {\rm {Tanh}}(\qW_{l}^{(i)} f_{l-1}^{(i)} + \qb_{l}^{(i)}), l=2,\dots,L
\end{equation}
where $f_1^{(i)}=\qz_0$, $\qW_{l}^{(i)}$ represents the weight matrix between layer $l$ and $l-1$ at iteration $i$. $\qb_{l}^{(i)}$ is the bias vector in layer $l$ at iteration $i$. In the beginning, each entry of $\qW_{l}^{(0)}$ and $\qb_{l}^{(0)}$ are initialized randomly following a uniform distribution with a range of $(\frac{-1}{\sqrt{p_l}},\frac{1}{\sqrt{p_l}})$ \cite{7410480}, where $p_l$ represents the number of neurons in layer $l$.
${\rm {Tanh}}$ is an activation function used after each layer.

After that, we use a stopping scheme in \cite{https://doi.org/10.48550/arxiv.2112.06074} to control the iterative process of D-DIP to avoid the overfitting problem due to the parameterization feature in the DIP. The stopping scheme is based on calculating the variance of the DNN output, given as 
\begin{align}\label{DIP_stop}
	\varsigma^{(i)} = \frac{1}{W} \sum_{j=i-W}^i \| \qx_{\rm {D-DIP}} ^{(j)} - \frac{1}{W} \sum_{j'=i-W}^i \qx_{\rm {D-DIP}} ^{(j')} \|^2, i\geq W,
\end{align}
where $	\varsigma^{(i)}$ is the variance value at iteration $i$. When $i<W$, the variance calculation is inactive. $W$ is a constant determined based on the experiments and should be smaller than the iterations needed for D-DIP to converge. As shown in Fig. \ref{fig:NN}, we compare $\varsigma^{(i)}$ with a threshold $\epsilon$. If $\varsigma^{(i)} < \epsilon$ the iterative process of D-DIP will stop, and the output of D-DIP $\qx_{\rm {D-DIP}}^{(I)}$ is then forwarded to BPIC as initial symbol estimates, i.e., $\hat{\qx}^{(0)} = \qx_{\rm {D-DIP}}^{(I)}$, where $I$ is the number of the last D-DIP iteration. Otherwise use mean square error (MSE) to calculate the loss shown as
\begin{equation}\label{loss}
    \mathcal{L}^{(i)}  = \frac{1}{2MN} \| \qH \qx_{\rm {D-DIP}}^{(i)} -  \qy \|^2.
\end{equation}
The DNN parameters that consist of weights $\qW_l^{(i)}$ and biases $\qb_l^{(i)}$ are then optimized by using Adam optimizer \cite{kingma2014adam} and the calculated loss in \eqref{loss}. The process is then repeated as shown in Fig. \ref{fig:NN}.

\section{Complexity Analysis}\label{Complexity}

In this section, we analyze the computational complexity of the proposed D-DIP-BPIC detector. As for the complexity of D-DIP, the computational complexity of fully-connected layers is matrix vector multiplications with a cost of $\mathcal{O} (M^2N^2I)$, where $I$ denotes the number of iterations needed for D-DIP. 
The computational complexity for different detection algorithms is shown in Table \ref{Tab:complexity}, where $T$ represents the iterations needed for the BPIC, UAMP, EP and BPICNet detectors.
For instance, for $M=12,N=7,T=10,I=50$, the complexity of D-DIP-BPIC is approximately 1.5 times lower than MMSE-BPIC, UAMP and BPICNet. The complexity of D-DIP-BPIC is approximately 15 times lower than EP. Thus our proposed detector has the lowest complexity compared to above high-performance detectors. 

Note that BPICNet has an extra complexity due to training requirements. BPICNet uses a large data set used for the training prior to deployment. For example, $b=5.12\times10^6$ is used in \cite{9900413}. 
%where $b$ represents the multiplications of the number of epochs, batch size and training samples needed in training BPICNet.Our D-DIP scheme avoids the costly training phase as compared to BPICNet.

\begin{figure}
	\centering
	\includegraphics[width=0.9\textwidth]{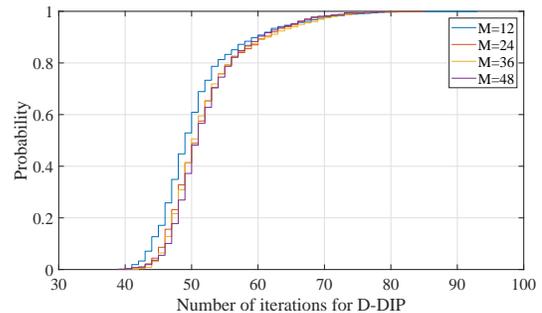}
	\caption{CDF of I}
	\label{fig:cdf}
\end{figure}

Fig. \ref{fig:cdf} shows the cumulative distribution function (CDF) of $I$ (i.e., the number of D-DIP iterations needed to satisfy the stopping scheme \eqref{DIP_stop}) for $M={12,24,36,48},N=7,l_{max}=M-1,k_{max}=3, SNR = 15dB$. The figure shows that the number of iterations required for D-DIP to converge, $I$, is not sensitive to the OTFS frame size (i.e., $M$ and $N$) which is a significant advantage.

\begin{table} \small  \addtolength{\tabcolsep}{-2.5pt}
	\begin{tabular}{| c| c| c|}
		\hline
		
		Detector 			&     \makecell	{Complexity order\\(Training)}	  & \makecell	{Complexity order\\(Deployment)} \\
		%\hline
		%\hline
		%MMSE\cite{singh2020low} 	& 		$\mathcal{O} (M^3N^3)$ 	\\
		%\hline
		%D-DIP 	& 		$\mathcal{O} (M^2N^2I)$ 	\\
		\hline
		%BPIC\cite{9569353} 		& 		$\mathcal{O}(M^2N^2T)$ 	 & 0.7056	\\
		MMSE-BPIC\cite{9569353} 	&Not required	& 		$\mathcal{O}(M^3N^3+M^2N^2T)$ 	\\
		\hline
		UAMP\cite{9293406} 		&Not required    & 		$\mathcal{O}(M^3N^3+M^2N^2T)$ 	 \\
		\hline
		EP\cite{long2021low} 	&Not required	& 		$\mathcal{O}(M^3N^3T)$ 	 	\\
            \hline
            BPICNet\cite{9900413} 	& \makecell	{$\mathcal{O}(b(M^3N^3+MN$\\$+M^2N^2T))$ }	& 		\makecell	{$\mathcal{O}(M^3N^3+MN$ \\ $+M^2N^2T)$ }	 	\\
		\hline
            D-DIP-BPIC 		&Not required & 		$\mathcal{O}(M^2N^2I + M^2N^2T)$ 	\\
            \hline
		%GEPNet\cite{9832663} & \makecell{$\mathcal{O}$  $((M^3N^3+M^2N^2K) T)$} & 60.68160	\\
		%\hline
	\end{tabular}  
	\caption{Computational complexity comparison}
	\label{Tab:complexity}
\end{table}

\section{Numerical results}\label{Results}

In this section, we evaluate the performance of our proposed detector by comparing its SER performance with those in MMSE-BPIC \cite{9569353}, UAMP \cite{9293406}, EP \cite{long2021low} and BPICNet \cite{9900413}. Here we use UAMP in \cite{9293406} instead, because the UAMP proposed in \cite{9492800} is not suitable for our system model as shown in \cite{9569353}. For the simulations, we set $N=7, l_{max}=M-1$, $\Delta f=15$kHz. The carrier frequency is set to $f_c=10$GHz. The $4$-QAM modulation is employed for the simulations, and we set $c=1/\sqrt{2}$ that is corresponding to the normalized power of constellations to normalize the DNN output. The same DNN parameters described in section \ref{D-DIP} (e.g., number of layers and number of neurons in each layer) are used in the DNN for all simulations. We use the Adam optimizer with a learning rate of $0.01$ to optimize the DNN parameters. The stopping criteria parameter for \eqref{DIP_stop}, $W$ is set to 30, and the threshold $\epsilon$ is set to 0.001. The number of iterations for the BPIC, UAMP, EP and BPICNet is set to $T=10$ to ensure convergence. 
For the training setting of BPICNet, we use the same setting in \cite{9900413}, where $M=12,N=7,l_{max}=11,k_{max}=3$ and 500 epochs are used during the training process, in each epoch, 40 batches of 256 samples were generated. $P \in \{6,\dots,12\}$ is randomly chosen and the values of SNR are uniformly distributed in a certain range, more details are shown in \cite{9900413}.

\begin{figure}
\centering
\subfigure[$M=12,P=6,k_{max}=3$]
{\includegraphics[width=0.48\textwidth]{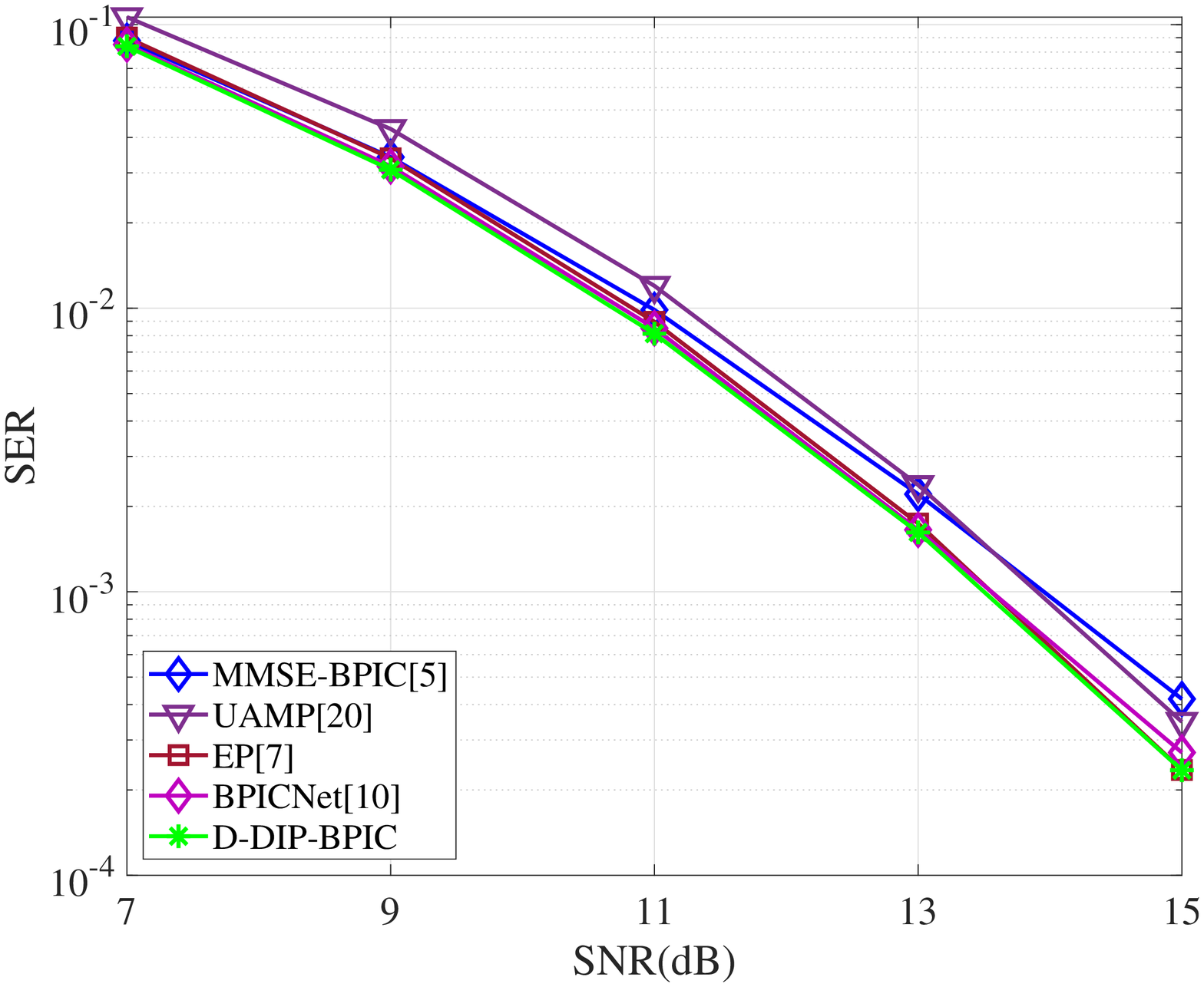}\hfill
\label{fig_snr}
}
\subfigure[$P=6, k_{max}=3$, SNR=15dB]
{\includegraphics[width=0.48\textwidth]{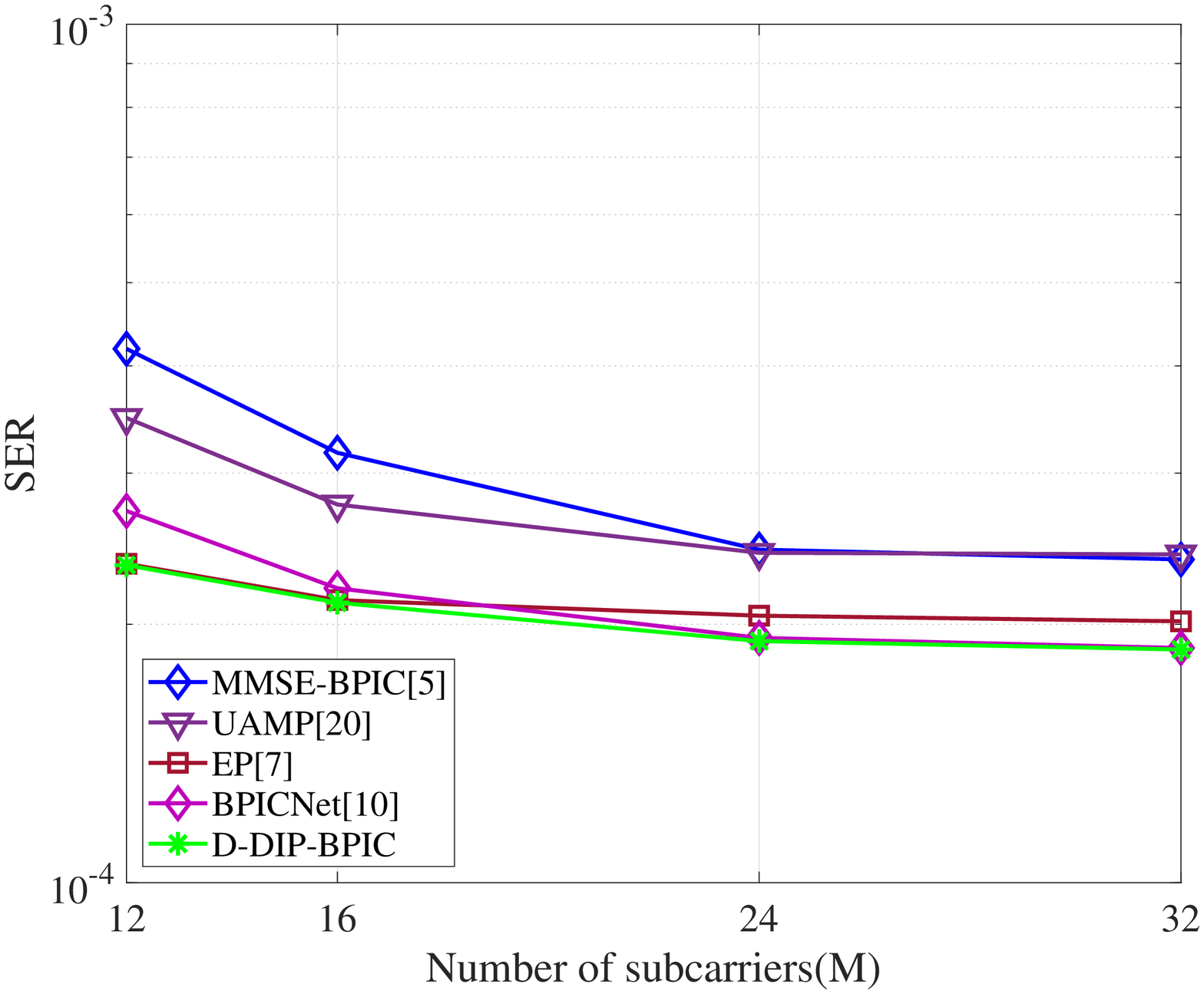}\hfill
\label{fig_scale}}

\subfigure[$M=12,k_{max}=3$, SNR=15dB]
{\includegraphics[width=0.48\textwidth]{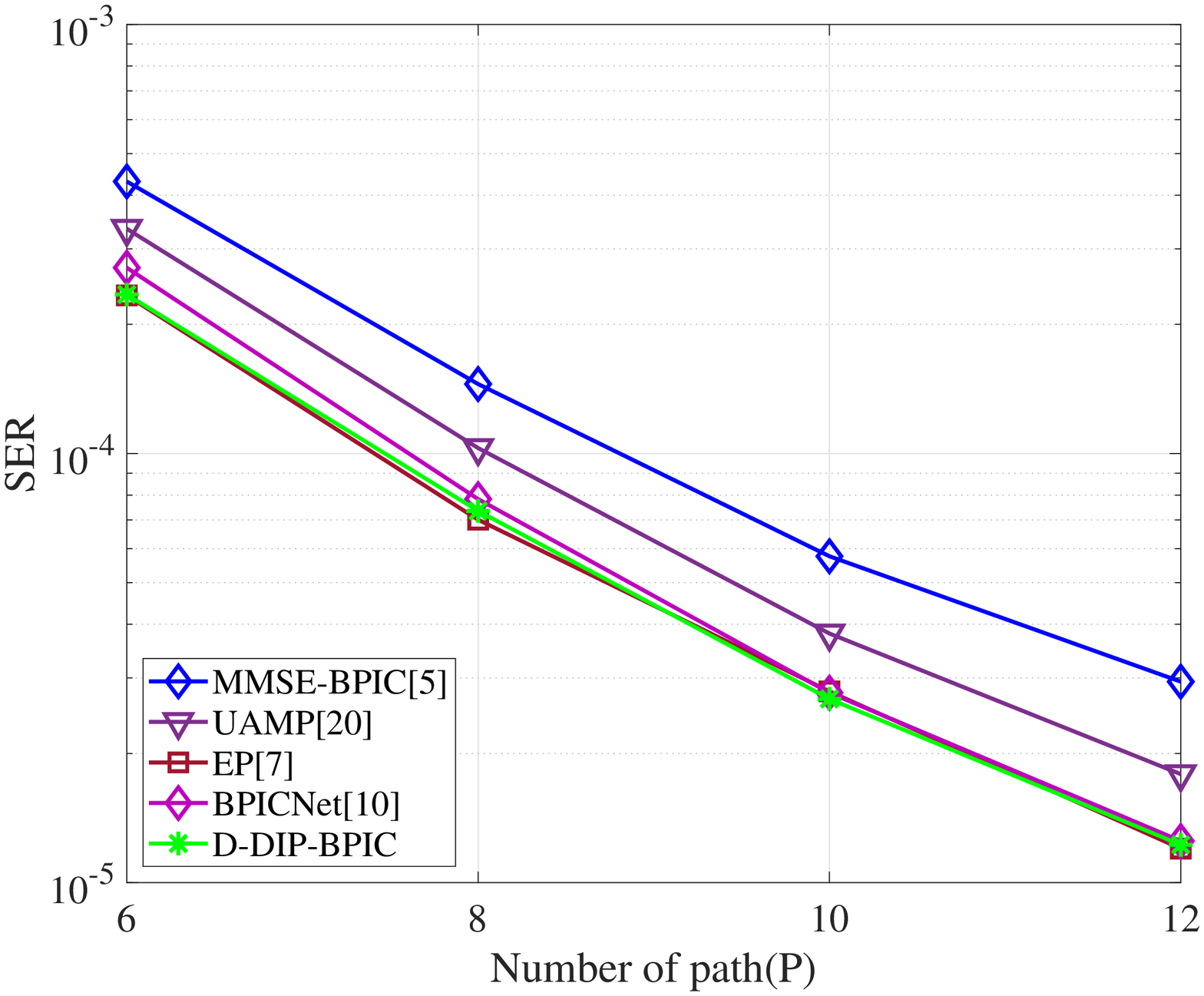}\hfill
\label{fig_path}
}
\subfigure[$M=12, P=12$, SNR=15dB]
{\includegraphics[width=0.48\textwidth]{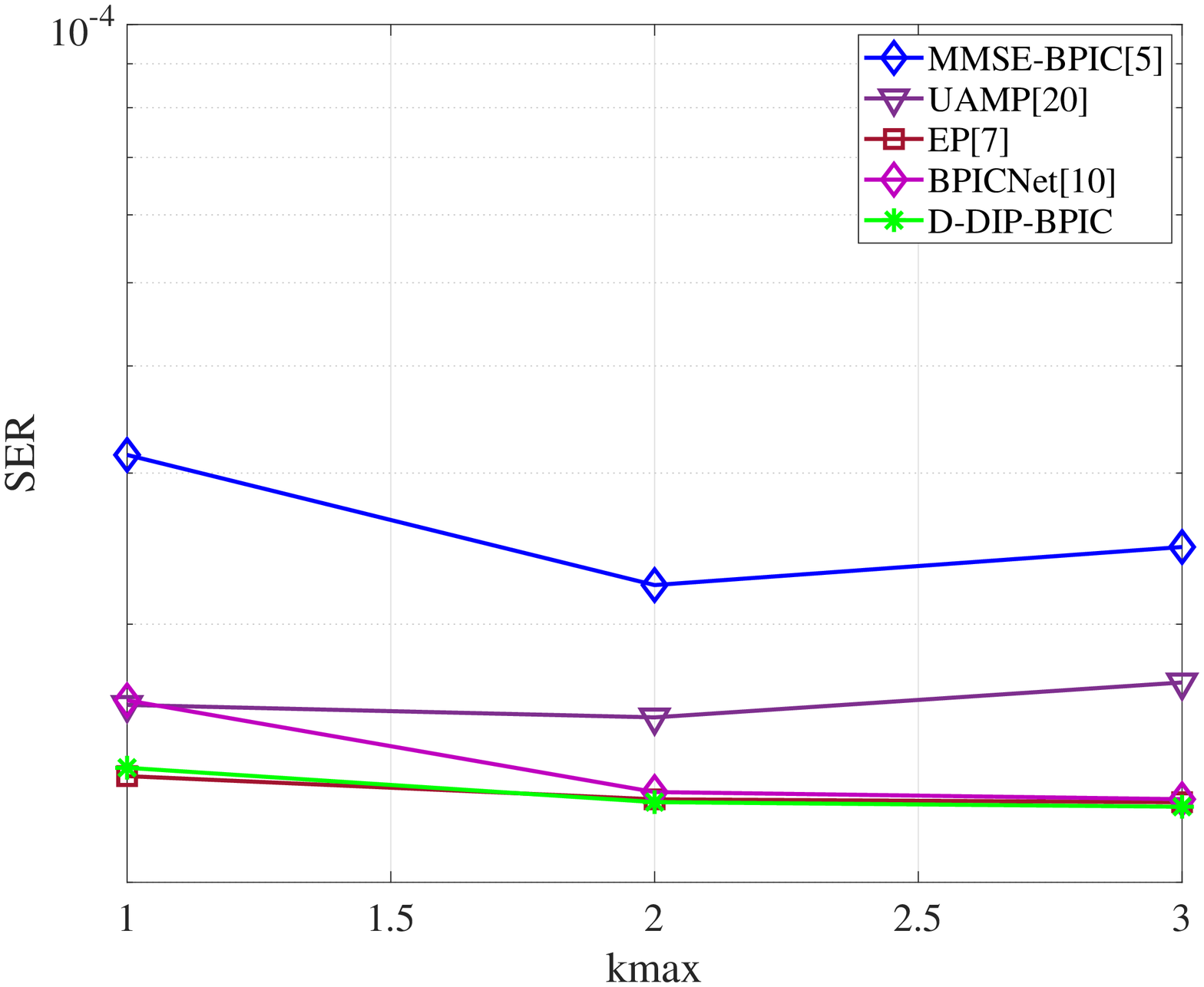}\hfill
\label{fig_kmax}}
\caption{SER performance}
\end{figure}

Fig. \ref{fig_snr} demonstrates that the proposed D-DIP-BPIC detector achieves around 0.5 dB performance gain over MMSE-BPIC and UAMP. In fact, its SER performance is very close to BPICNet and EP.
Fig. \ref{fig_scale} evaluates the scalability of our proposed D-DIP-BPIC detector. As we increase the OTFS frame size (i.e., number of subcarriers), D-DIP-BPIC remains the outperformance over MMSE-BPIC and UAMP and achieves a close to BPICNet and EP performance.
Fig. \ref{fig_path} shows that when the number of paths (e.g., mobile reflectors) increases, the D-DIP-BPIC detector still can achieve close to BPICNet and EP performance and outperform others.
As shown in Fig. \ref{fig_kmax}, it is obvious that the performance of the BPICNet detector degrades in the case of $k_{max}=1$ as compared to $k_{max}=2,3$ as the fidelity of training data is compromised while our D-DIP-BPIC still retains its benefit.

\section{Conclusion}
We proposed an untrained neural network based OTFS detector that can achieve excellent performance compared to state-of-the-art OTFS detectors. Our simulation results showed that the proposed D-DIP-BPIC detector achieves a 0.5 dB SER performance improvement over MMSE-BPIC, and achieve a close to EP SER performance with much lower complexity.

%
%\section*{Acknowledgment}
%This research was supported by the research training program stipend from The University of Sydney. The work of Branka Vucetic was supported in part by the Australian Research Council Laureate Fellowship grant number FL160100032.

{\renewcommand{\baselinestretch}{1.1}
	\begin{footnotesize}
		\bibliographystyle{IEEEtran}
		\bibliography{IEEEabrv,myBib}
\end{footnotesize}}

\end{document}